\documentclass[aps,prd,twocolumn,showpacs,superscriptaddress,groupedaddress]{revtex4}
\usepackage{graphicx}
\usepackage{dcolumn}
\usepackage{bm}
\usepackage{textcomp} 
\usepackage{amstext} 
\usepackage[utf8]{inputenc}
\usepackage{graphicx}
\usepackage{epsfig}
\usepackage{mathtools, nccmath}
\usepackage{environ}
\usepackage{tabularx}
\usepackage[usenames]{color}
\NewEnviron{myequation}{%
\begin{equation}
\scalebox{0.85}{$\BODY$}
\end{equation}}
\NewEnviron{myequation2}{%
\begin{equation}
\scalebox{0.74}{$\BODY$}
\end{equation}}
\usepackage{orcidlink}
\hypersetup{pdfstartview=FitH, linkcolor=blue,urlcolor=blue, colorlinks=true}

\begin{document}

\hspace{5.2in} \mbox{ILD-PHYS-2022-001}

\vspace*{0.1cm}

\title{Measurement of {\boldmath ${\cal B}r(H \to Z\gamma)$} at the 250 GeV ILC}


\author{E.~Antonov\,\orcidlink{0000-0001-7665-6493} and A.~Drutskoy\,\orcidlink{0000-0003-4524-0422} \vspace*{0.5cm}}

\begin{abstract}
\vspace*{0.01cm}
The $e^+e^- \to HZ$ process with the subsequent decay of the Higgs boson $H \to Z\gamma$ is studied, where both $Z$ bosons are reconstructed in the final states with two jets. The analysis is performed using Monte Carlo data samples obtained with detailed ILD detector simulation assuming an integrated luminosity of 2~ab$^{-1}$, beam polarizations of ${\cal{P}}_{e^-e^+} = (-0.8, +0.3)$, and center-of-mass energy of \mbox{$\sqrt{s}$ = 250~GeV} for the electron-positron International Linear Collider being currently designed. The analysis is repeated for the case of two 0.9~ab$^{-1}$ data samples with polarizations ${\cal{P}}_{e^-e^+} = (\mp0.8, \pm0.3)$. Contributions of the potential background processes are studied using all available ILD MC event samples. The largest background comes from the $e^+e^- \to W^+W^-$ process supplemented by an energetic photon produced by initial state radiation. To suppress this background we require that at least one of the $Z$ bosons decays to $b$-jets. To reduce the jet reconstruction uncertainties the $M_{\Delta} = M(jj\gamma) - M(jj) + M(Z_{\rm nom})$ variable is used, where $M(Z_{\rm nom})$ = 91.2~GeV. The $M_{\Delta}$ distributions are obtained for the studied signal and backgrounds to estimate the expected accuracy of the ${\cal B}r(H \to Z\gamma)$ measurement. The accuracy is 22$\%$ for the option of the single polarization sample described above and deteriorate to 24$\%$ in case of the sample with two polarizations. The proposed method can be applied at any future $e^+e^-$ collider.
\end{abstract}
\smallskip

\maketitle

\section{\label{sec:intro}Introduction}

The discovery of the Higgs boson by the ATLAS and CMS collaborations~\cite{atlas,cms}~in 2012 initiated measurements of Higgs boson parameters with increasing accuracy. Although many of these parameters can be precisely measured at the LHC, the most of them can be more accurately determined at future $e^+e^-$ colliders. In particular, decay channel $H \to Z\gamma$ is well suited for lepton colliders. The ATLAS~\cite{atlasrun2} and CMS~\cite{cms2022} collaborations estimated the expected statistical significance for the signal from the Standard model (SM) Higgs boson to be 1.2$\sigma$ based on full datasets accumulated at $\sqrt{s}$ = 13 TeV. ATLAS Collaboration expects to reach 19\% precision in the measurement of the product of the Higgs boson production cross section and the branching fraction with 3000 fb$^{-1}$ dataset to be accumulated at the High Luminosity LHC at 14 TeV in the future~\cite{lhc}.

\begin{figure}[!ht]\label{fig:Diag} 
\centering
\includegraphics[scale=0.16]{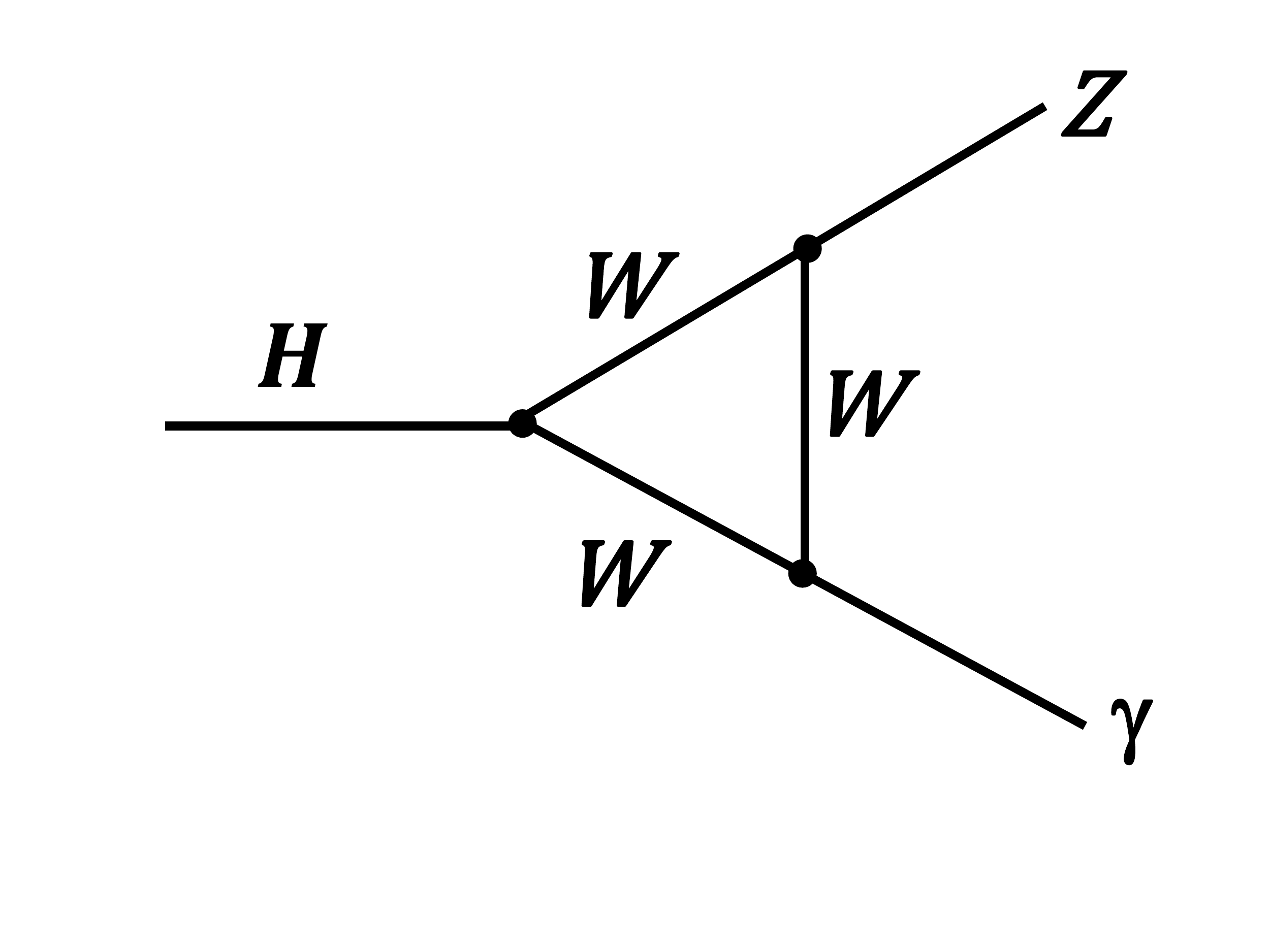}\includegraphics[scale=0.16]{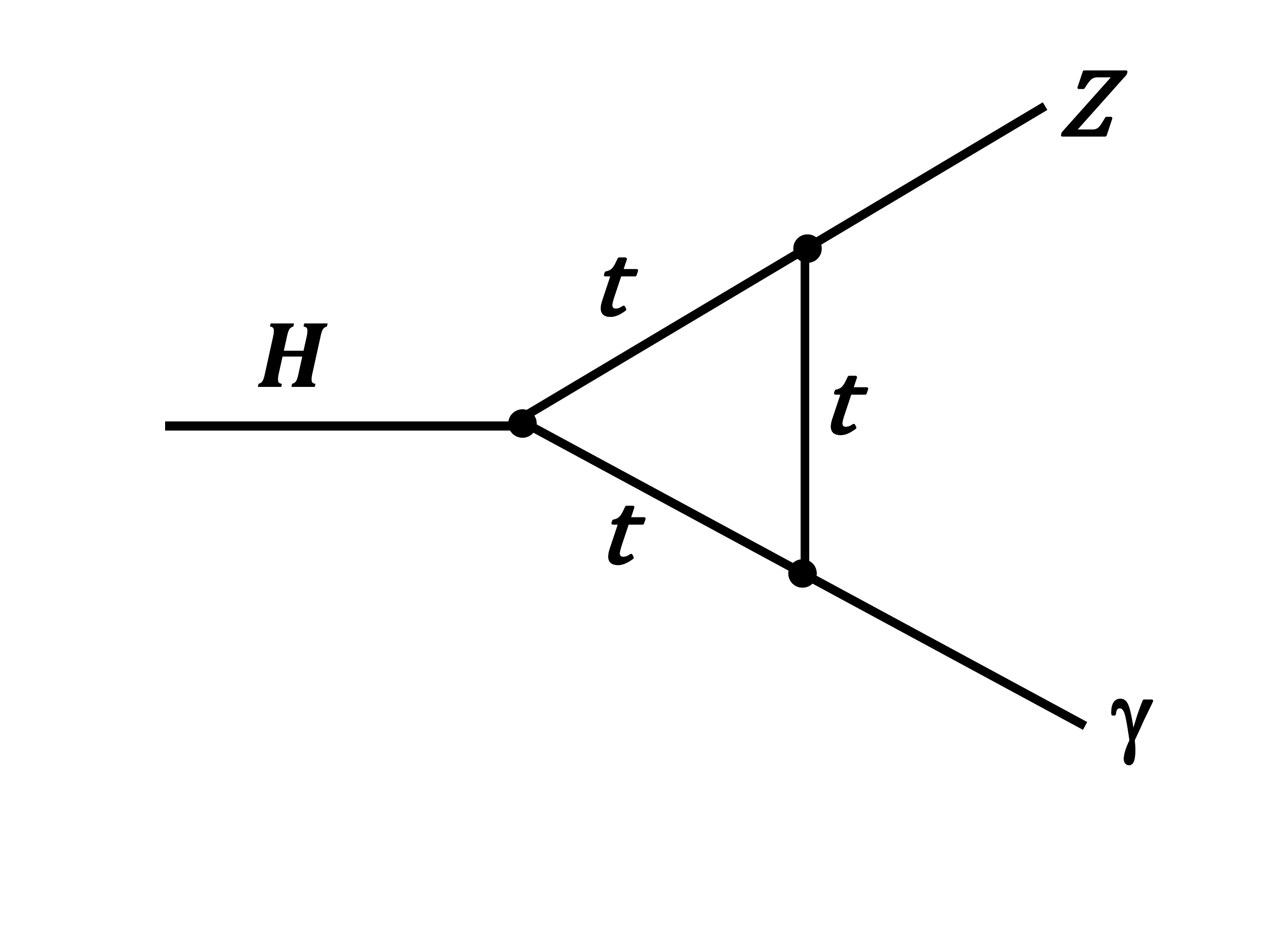}
\caption{The loop diagrams of the decay of the Higgs boson $H \to Z\gamma$. The channel with the $W$ loop is expected to give a dominant contribution of about 85-90$\%$.}
\end{figure}

The process $H \to Z\gamma$ is described within the SM by the loop diagrams (\hyperref[fig:Diag]{Fig.~1}) with charged particles inside the loop. Potentially heavy charged particles predicted within any extension of the Standard Model can also contribute to the process in a similar manner ~\cite{hzzm}. Although the $H \to Z\gamma$ decay branching fraction depends weakly on the mass of the $W$ boson~\cite{wmass}, this effect is well inside uncertainties obtained in this analysis.

Using Monte Carlo (MC) simulation of the proposed experiment at the CEPC collider being designed in China, the accuracy of the ${\cal B}r(H \to Z\gamma)$ measurement was estimated assuming a data sample of 5.6 ab$^{-1}$~\cite{cepc}. In that analysis the $e^+e^- \to Z H(Z\gamma)$ process was reconstructed requiring that one of the $Z$ bosons decays in the $Z \to \nu\bar{\nu}$ channel and the other in the $Z \to jj$ channel. These two possible combinations of the respective final states were studied together and the accuracy of 13$\%$ was obtained for the ${\cal B}r(H \to Z\gamma)$ measurement.

In this analysis we reconstruct both $Z$ bosons using hadronic jets to increase the statistics. Below we denote the directly produced $Z$ boson as $Z_1$ to separate it from the $Z$ boson produced in the Higgs decay. There are a number of background processes with high cross sections that have a similar final configuration including four jets and a high energy photon. The largest background comes from the $e^+e^- \to W^+W^-\gamma$ process where the photon is produced by initial state radiation (ISR) from the beams. To suppress this background we require that at least one $Z$ boson decays to $b\bar{b}$ jets. The $b$-jet tagging technique provides a high efficiency for the signal and a strong $W^+W^-$ channel suppression. Finally we analyse the process:

\begin{eqnarray}\label{eq:channel1}
  e^+e^- & \to Z_1(\lowercase{q \bar{q}})\,H,  \ \ \ \ \ \  H \to Z(\lowercase{q \bar{q}}) \gamma,
\end{eqnarray}

\noindent
where either of the two quark pairs can be the $b\bar{b}$ pair. For simplicity, all jet flavours are analyzed together, future improvements should include flavour-specific corrections ~\cite{flavor}.

\section{\label{sec:sample}MC samples and analysis tools}

All available official MC data samples produced by the ILD Collaboration are used. All processes are generated using the Whizard 2.8.5 package~\cite{whizard} with the LCIO~\cite{lcio} output format; hadronization is performed by Pythia6~\cite{pythia6}. The detailed simulation of the ILD (The International Large Detector) detector is performed using the \texttt{ILD\char`_l5\char`_o1\char`_v02} model from the ILCSoft toolkit~\cite{ilcsoft} \texttt{v02-00-02} using the DD4HEP~\cite{dd4hep} software package. Finally, the events are reconstructed with Marlin~\cite{marlin}.

The official MC samples are generated assuming four possible combinations with 100$\,\%$ beam polarization, ${\cal{P}}_{e^-e^+} = (\pm\,1.0, \pm\,1.0)$, and 250 GeV center-of-mass energy. Initial state radiation and beam radiation processes are properly included at the generation level. Low-$p_t$ hadrons produced from high-rate $\gamma\gamma$-induced processes as well as $e^+e^-$ pairs from beamstrahlung are overlaid on the simulated events before reconstruction. The MC samples contain information about all particles in an event. In particular, the MCParticles~\cite{mcp} and \mbox{PandoraPFOs} (the Particle Flow Objects reconstructed by \mbox{PandoraPFA}~\cite{pfo}) are important for the studies presented. Table~\ref{tab:tab1} shows the basic information for the MC samples most important for this analysis.

\renewcommand{\arraystretch}{1.2}
\begin{table*}[htb]
\caption{The basic information for the MC samples for the signal and the backgrounds with significant contributions. The given cross sections are corrected for the decay branching fractions indicated in the first column.}
\label{tab:tab1}
\begin{center}
\begin{tabular}
{@{}l@{\hspace{0.5cm}} @{\hspace{0.6cm}}c@{\hspace{0.6cm}} @{\hspace{0.6cm}}c@{\hspace{0.6cm}} @{\hspace{0.6cm}}c@{\hspace{0.6cm}} @{\hspace{0.5cm}}c@{\hspace{0.6cm}} @{\hspace{0.6cm}}c@{\hspace{0.6cm}} @{\hspace{0.6cm}}c@{\hspace{0.00001cm}}}
\hline \hline
Process & \multicolumn{2}{c}{Integrated luminosity, ab$^{-1}$} & \multicolumn{2}{c}{Cross section, fb} & \multicolumn{2}{c}{Number of events} \\
\hline
$e^-$/$e^+$ polarization & eLpR & eRpL & eLpR & eRpL & eLpR & eRpL \\
\hline
 & \multicolumn{6}{c} {\small{Signal samples}}\\
\hline
\small{$q\bar{q}H(Z\gamma)$} & \small{191} & \small{298} & \small{0.52} & \small{0.34} & \small{1$\cdot$10$^5$} & \small{1$\cdot$10$^5$} \\
\hline \hline
 & \multicolumn{6}{c} {\small{Background samples}}\\
\hline
\small{$q\bar{q}$} & \small{5.00} & \small{5.00} & \small{128$\cdot$10$^3$} & \small{70.4$\cdot$10$^3$} & \small{6.40$\cdot$10$^8$} & \small{3.52$\cdot$10$^8$} \\
\small{$W(q\bar{q})W(q\bar{q})$} & \small{5.00} & \small{5.12} & \small{14.8$\cdot$10$^3$} & \small{225} & \small{7$\cdot$10$^7$} & \small{7$\cdot$10$^5$} \\
\small{$Z(q\bar{q})Z(q\bar{q})$} & \small{5.05} & \small{5.11} & \small{1.41$\cdot$10$^3$} & \small{607} & \small{7$\cdot$10$^6$} & \small{3$\cdot$10$^6$} \\
\small{$Z/W(q\bar{q})Z/W(q\bar{q})$} & \small{5.00} & \small{5.32} & \small{12.4$\cdot$10$^3$} & \small{226} & \small{6$\cdot$10$^7$} & \small{10$^6$} \\
\small{$Z(q\bar{q})Z(\mu^+\mu^-/\tau^+\tau^-)$} & \small{5.01} & \small{5.14} & \small{838} & \small{467} & \small{4$\cdot$10$^6$} & \small{2$\cdot$10$^6$} \\
\small{$q\bar{q}H(b\bar{b})$} & \small{0.50} & \small{0.78} & \small{199} & \small{128} & \small{$10^5$} & \small{$10^5$} \\
\small{$q\bar{q}H(\tau^+\tau^-)$} & \small{23.2} & \small{36.3} & \small{21.5} & \small{13.8} & \small{5$\cdot$10$^5$} & \small{5$\cdot$10$^5$}\\
\small{$q\bar{q}H(W^+W^-)$} & \small{6.81} & \small{10.6} & \small{73.4} & \small{47.0} & \small{5$\cdot$10$^5$} & \small{5$\cdot$10$^5$} \\
\small{$q\bar{q}H(ZZ)$} & \small{55.6} & \small{86.9} & \small{8.99} & \small{5.75} & \small{5$\cdot$10$^5$} & \small{5$\cdot$10$^5$} \\
\small{$\tau^+\tau^-H($all$)$} & \small{7.45} & \small{11.6} & \small{67.1} & \small{42.9} & \small{5$\cdot$10$^5$} & \small{5$\cdot$10$^5$} \\
\hline \hline
\end{tabular}
\end{center}
\end{table*}
\linespread{1.3}

The separation of isolated photons, jet reconstruction via FastJet~\cite{fj}, and $b$-jet tagging are handled by additional Marlin processors as described in the following sections.

To get the expected number of signal or background events with ${\cal{P}}_{e^-e^+} = (-0.8, +0.3)$ polarization and the integrated luminosity 2 ab$^{-1}$, we apply a weight factor to each event from the MC samples. The sample nominal integrated luminosities $\mathcal{L}$ are given in Table~\ref{tab:tab1}. The weight factor $W_{LR/RL}$ depends on the target polarisations and the beam chiralities for which the original MC events have been generated, as well as on the ratio of the target luminosity to the generated luminosity. For the target values of $P_{e^-e^+}=(-0.8,+0.3)$ and 2\,ab$^{-1}$, the weights are:

\begin{fleqn}[\parindent]
\begin{equation} \label{eq:wfactor}
W_{LR/RL} = \left[ \frac{(1\pm0.8)}{2}\cdot\frac{(1\pm0.3)}{2} \right] \cdot \frac {\rm 2~ab^{-1}} {\mathcal{L}}
\end{equation} 
\end{fleqn}

The numbers of the MC generated events before weighting are significantly larger than the numbers of expected events obtained after weighting.

\section{\label{sec:selection} Event preselection and initial analysis}

The signal MC samples are preselected requiring only specific process and decay chains. All following selections are applied using the information on the reconstruction level.

The first step of the event selection is identification of the isolated photon candidate. The \texttt{IsolatedPhotonTagging} processor is applied for this goal. This processor finds isolated high energy photons in events using a double-cone method and TMVA~\cite{tmva} machine learning algorithms. We used the default set of parameters and weights included in this processor.

\begin{figure}[!ht]\label{fig:Egamma} 
\centering
\includegraphics[scale=0.4]{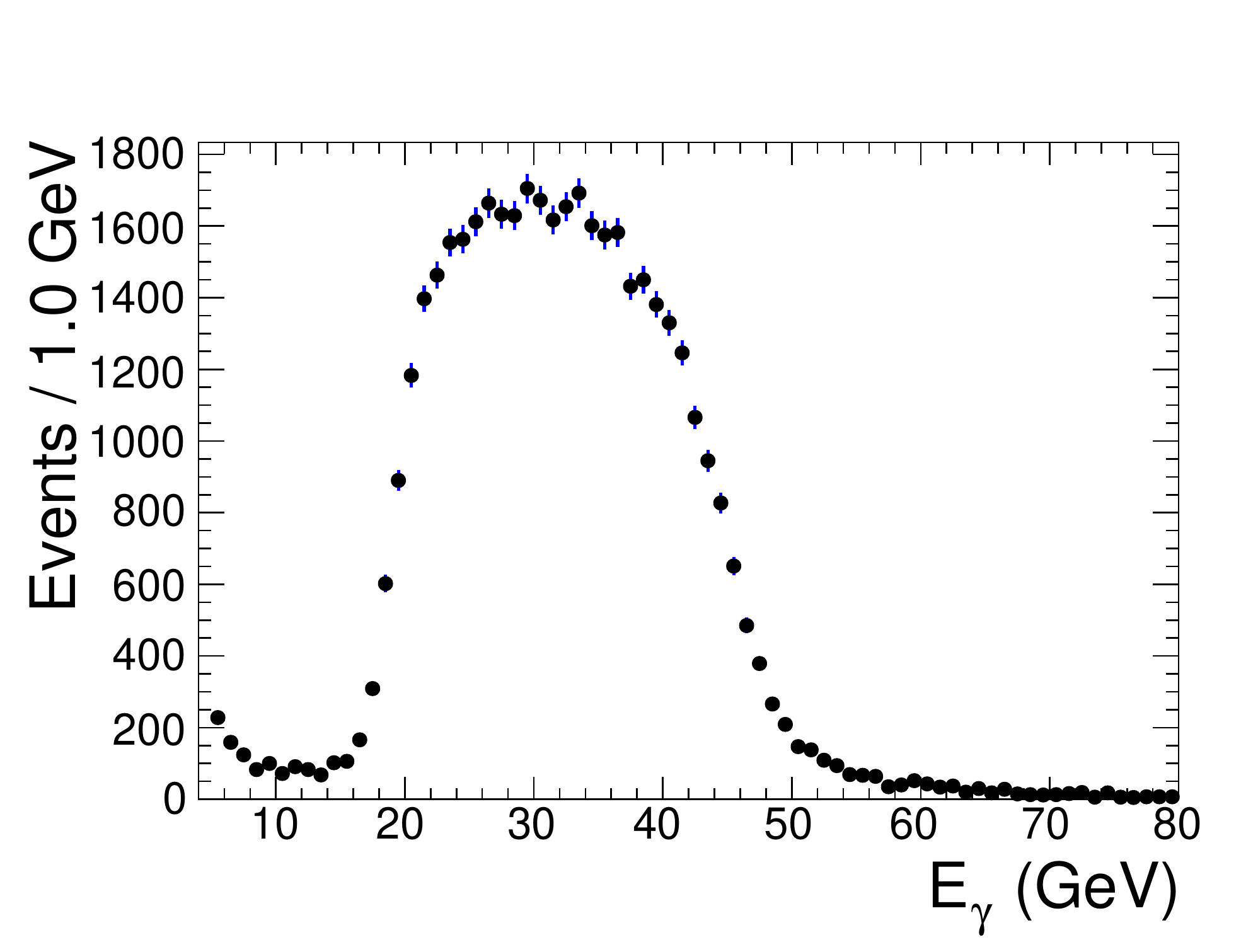}
\caption{The energy distribution of identified photons produced in the process $e^+e^- \to HZ$, with $H \to Z\gamma$.}
\end{figure}

Two variables are used to suppress the background contributions related to ISR photons: the photon energy $E_{\gamma}$, and the angle between the photon and beam directions ${\rm cos} \theta_{\gamma\mathchar`-beam}$. The ISR photons are mostly located in the regions close to the beam directions and at low energy. The signal photons have a flat angular distribution in ${\rm cos} \theta_{\gamma\mathchar`-beam}$ and concentrate in the energy region shown in \hyperref[fig:Egamma]{Fig. 2}. First, we choose the isolated photon with $ E_{\gamma} > 5$~GeV, which has the maximum energy over all photons in the event.

\begin{figure}[!ht]\label{fig:PgCosg} 
\centering
\includegraphics[scale=0.4]{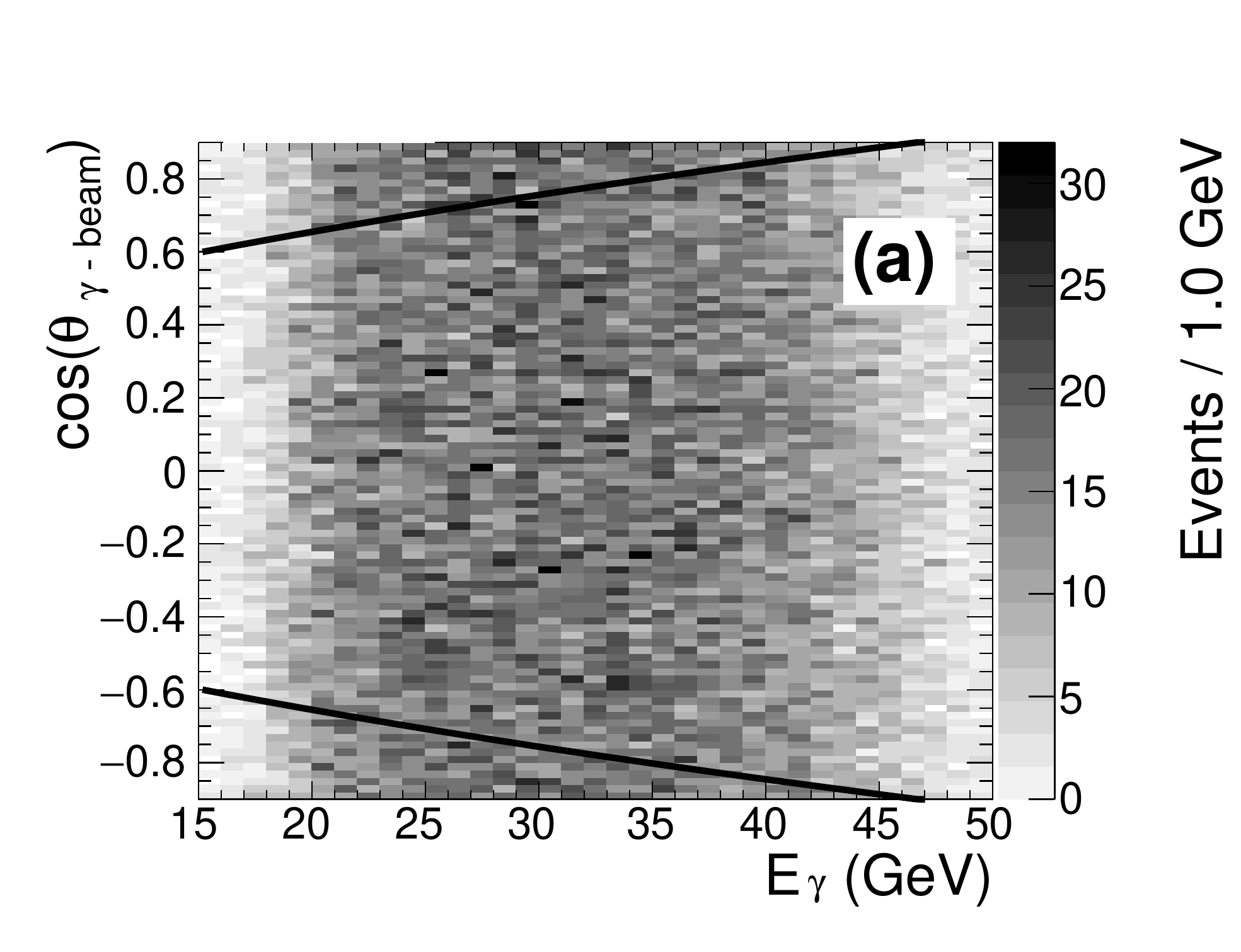}\vspace{-0.3cm}
\includegraphics[scale=0.4]{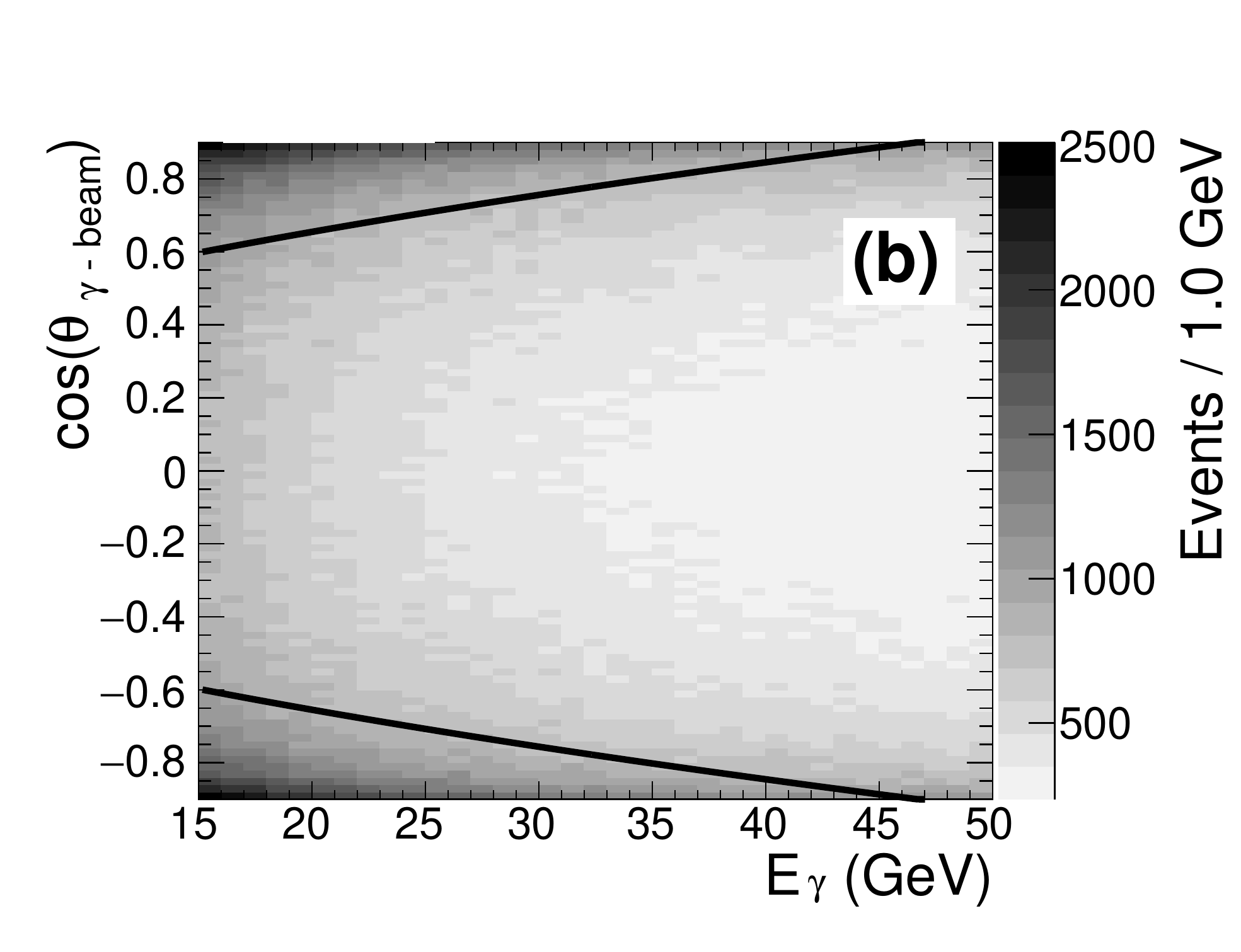}\vspace{-0.2cm}
\caption{Two-dimensional distributions ${\rm cos} \theta_{\gamma\mathchar`-beam}$ vs $E_{\gamma}$ for the signal (a)
and the sum of significant backgrounds (b). The curves indicate the cut explained in the text.}
\end{figure}

We applied cuts on the energy and polar angle of the photons, $E(\gamma) = [18, 50]$~GeV and $|{\rm cos} \theta_{\gamma\mathchar`-beam}| < 0.95$, respectively. The photons from the background process $e^+e^- \to Z\gamma$ have the momentum 108~GeV and are removed by these cuts. To further suppress ISR photons, the two-dimensional cut  $E_{\gamma} - 70 \cdot {\rm cos}^2 \theta_{\gamma\mathchar`-beam} > -10$~GeV is applied. The cut is shown in \hyperref[fig:PgCosg]{Fig. 3}, where the two-dimensional distributions ${\rm cos} \theta_{\gamma\mathchar`-beam}$ vs $E_{\gamma}$ are given for the signal (a) and all background contributions (b). 

The most dangerous background sources are the $e^+e^- \to W^+W^-\gamma_{\rm \,ISR}$ and $e^+e^- \to ZZ\gamma_{\rm \,ISR}$ processes, which include four jets and an ISR photon in the final states. The background $e^+e^- \to W^+W^-\gamma_{\rm \,ISR}$ for eLpR polarization has a large cross section and must be strongly suppressed. For this goal we require that at least one jet is tagged as a $b$-jet. To reconstruct jets and to tag $b$-jets in event, we use the FastJet software package with the Valencia~\cite{vlc} algorithm, which was specially developed for jet reconstruction at electron-positron colliders. We select this algorithm for its high efficiency of jet reconstruction near the beam direction. Three parameters should be adjusted in the Valencia algorithm, the generalized jet cone radius $R$, and the $\beta$ and $\gamma$ parameters, which are used to control the clustering order and the background resilience. We set $\beta$ to~$1.0$, $\gamma$ to~$0.5$ and $R$ to~$1.5$ and force all particles except the identified photon to form four jets.

The $b$-jet tagging MVA likelihood is calculated by the LCFI+ (Linear Collider Flavor Identification) algorithm ~\cite{lcfiplus}. The jets with $b$ tagging MVA likelihood value larger than 90$\,\%$ are taken as positively identified. Only events with at least one positively identified $b$-jet are selected for the following analysis. To obtain the efficiency of identification, the ratio of the number of events with identified $b$-jet to the number of events containing $b$-quarks at the generator level is calculated. This identification efficiency is $\sim$87$\,\%$ for all beam polarizations. The branching fractions of the $Z$ boson decay to any flavour hadron jets and the $b\bar{b}$ jets are equal to $69.91\pm0.06 \%$ and $15.12\pm0.05\,\%$~\cite{pdg}, respectively. Therefore the efficiency to tag positively at least one \mbox{$b$-jet} over all 4-jet signal events is $\sim\,$34$\,\%$. The probability to tag a $b$-jet for events with 4 jets in the case of the absence of $b$-quarks on the MC generator level is 0.85$\,\%$.

The product of the cross section and the branching fraction discussed above can be measured experimentally using the formula: 

\begin{fleqn}[\parindent]
\begin{equation}\label{eq:sigmaa}
\begin{split}
\sigma(e^+&e^- \to H Z_1) \times Br( H \to Z \gamma) = \\
& {N_{\rm sig}} / ( \mathcal{L}_{\rm int} \cdot \epsilon \cdot Br(Z_1) \cdot Br(Z) )
\end{split}
\end{equation}
\end{fleqn}

\noindent
where $N_{\rm sig}$ is the number of signal events measured in a specific channel, and $\mathcal{L}_{\rm int}$ is the integrated luminosity of a used data sample. The selection efficiency is denoted by $\epsilon$ and the relevant decay branching fractions of the $Z$ boson decays taken from PDG (Particle Data Group)~\cite{pdg} are denoted as $Br(Z_1)$ and $Br(Z)$.

To obtain a better resolution, the number of Higgs boson signal events is obtained by fitting the $M_{\Delta}$ distribution, where the $M_{\Delta}$ is calculated from the following formula:

\begin{fleqn}[\parindent]
\begin{equation}\label{eq: Mass Difference}
M_{\Delta} = M(jj\gamma) - M(jj) + M(Z_{\rm nom})
\end{equation}
\end{fleqn}

\noindent where $M(Z_{\rm nom}) = 91.2$~GeV. This formula results in a narrower Higgs boson mass peak, because uncertainties of the jet reconstruction are mostly canceled in the mass difference.

\section{\label{sec:final} Results}

The final state of the signal channel includes one photon and four jets. To form the $Z_1$ and $Z$ bosons from these four jets we calculate a $\chi^2$ for six possible two-jet combinations using respective masses and momenta:

\begin{fleqn}[\parindent]
\begin{myequation}\label{eq:chi_sq}
\begin{multlined}
{\chi}^2 = \frac{(M(Z_1)-M(Z_{\rm nom}))^2}{{{\sigma}{^2}_{M_{Z_1}}}} + \frac{(M(Z)-M(Z_{\rm nom}))^2}{{{\sigma}{^2}_{M_Z}}} \\
+ \frac{(P(Z_1)-\overline{P}(Z_{1}))^2}{{{\sigma}{^2}_{P_{Z_1}}}} + \frac{(P(Z+\gamma)-\overline{P}(Z_{1}))^2}{{{\sigma}{^2}_{P_{Z\gamma}}}}
\end{multlined}
\end{myequation}
\end{fleqn}
\noindent where $\overline{P}(Z_{1}) = 60.0$~GeV/$c$ is the mean $Z_{1}$ momentum in the $e^+e^- \to HZ_1$ process at the 250~GeV center-of-mass energy. The $\sigma$ parameters with the values ${\sigma}_{M_{Z_1}} = 14.2$~GeV, ${\sigma}_{M_{Z}} = 14.3$~GeV, ${\sigma}_{P_{Z_1}} = 7.1$~GeV, ${\sigma}_{M_{Z\gamma}} = 7.7$~GeV are the mean effective
widths of the corresponding mass or momentum distributions on the reconstruction level. The combination with the minimal 
$\chi^2$ is selected. Only events with the value $\chi^2 <\,$15 are kept for the following analysis.

The list of considered backgrounds includes the processes $e^+e^- \to q\bar{q}$, $e^+e^- \to Z(q_1\bar{q_1})Z(q_2\bar{q_2 })$ and $ e^+e^- \to \tau^+\tau^-H$. Background contributions come also from the $e^+e^- \to W^+(q_1\bar{q_1})W^-(q_2\bar{q_2})$ and $e^+e^- \to Z(q \bar{q})H(\tau^+\tau^-)$ processes. The $e^+e^- \to q\bar{q}$ process can be wrongly identified as a four-jet process and gives a large background contribution due to the huge production cross section. Some background processes are already partially suppressed due to the requirement of the $b$-jet tag.

We applied a set of additional cuts to further suppress backgrounds. The cuts are applied on the masses of the $Z_1$ and $Z$ bosons $M(Z_1 / Z) > 60$~GeV. Transverse momentum of the total system $P_t(jjjj\gamma) < 10$~GeV and the energy of the total system $E(jjjj\gamma) < 270$~GeV are bounded. A requirement is imposed on the value of the helicity angle ${\rm cos} \theta_h$ in the interval [$-0.95$, $+0.9$]. The helicity angle is calculated as the angle between the directions of the hadron jet with the higher momentum in the Higgs boson decay and the reconstructed Higgs boson $H(jj\gamma)$. This cut preserves most of the signal events but suppresses the main backgrounds with a high efficiency. An additional suppression of the background is provided by the cut on the angle between the $Z_1$ and $Z$ bosons ${\rm cos} \theta_{Z_1Z} > -0.95$. Possible purely leptonic backgrounds are rejected by the requirement on the total number of reconstructed objects in the event $N_{PFOs} > 60$. The numbers of signal and background events before and after cuts corresponding to the integrated luminosity 2 ab$^{-1}$ and polarization ${\cal{P}}_{e^-e^+} = (-0.8, +0.3)$ are listed in Table~\ref{tab:tab2} and Table~\ref{tab:tab3}, respectively.

\renewcommand{\arraystretch}{1.2}
\begin{table}[htb]
\caption{The numbers of signal events before and after cuts.
The percentage of the number of events from the previous step is indicated in brackets.}

\begin{center}
\label{tab:tab2}
\begin{tabular}
{@{\hspace{0.0cm}}l@{\hspace{0.3cm}}c@{\hspace{0.2cm}}c@{\hspace{0.2cm}} @{\hspace{0.01cm}}c@{\hspace{0.01cm}}}
\hline \hline
$e^+e^- \to Z_1(jj)Z(jj)\gamma$ & eLpR & eRpL \\
\hline
MC events & 70100 & 69786 \\
Weight factors & $6.1\cdot10^{-3}$ & $2.4\cdot10^{-4}$ \\
Weighted MC events & 430.5 & 16.4 \\
Photon tagging & 388.9 (90.3$\%$) & 14.8 (90.4$\%$)  \\
$b$-tagging & 131.5 (33.8$\%$) & 5.0 (34.0$\%$) \\
Weighted events after all cuts & 58.0 (44.1$\%$) & 2.0 (39.0$\%$)  \\
\hline
\hline
\end{tabular}
\end{center}
\end{table}

\begin{table*}[htb]
\caption{The numbers of MC events before and after cuts for significant backgrounds. The numbers of MC events before weighting are given in the nominal MC events column.}
\begin{center}
\label{tab:tab3}
\begin{tabular}
{@{}l@{\hspace{0.1cm}} @{\hspace{0.1cm}}c@{\hspace{0.1cm}} @{\hspace{0.1cm}}c@{\hspace{0.1cm}} @{\hspace{0.1cm}}c@{\hspace{0.1cm}} @{\hspace{0.1cm}}c@{\hspace{0.1cm}} @{\hspace{0.1cm}}c@{\hspace{0.1cm}} @{\hspace{0.1cm}}c@{\hspace{0.1cm}} @{\hspace{0.1cm}}c@{\hspace{0.1cm}} @{\hspace{0.1cm}}c@{\hspace{0.1cm}} @{\hspace{0.1cm}}c@{\hspace{0.1cm}} @{\hspace{0.1cm}}c@{\hspace{0.1cm}} @{\hspace{0.1cm}}c@{\hspace{0.1cm}} @{\hspace{0.1cm}}c@{\hspace{0.1cm}} @{\hspace{0.1cm}}@{}}
\hline \hline
Process & \multicolumn{2}{c}{\small{Nominal}} & \multicolumn{2}{c}{\small{Weight factors}} & \multicolumn{2}{c}{\small{Weighted}} & \multicolumn{2}{c}{\small{$\gamma$-tagging}} & \multicolumn{2}{c}{\small{$b$-tagging}} & \multicolumn{2}{c}{\small{After all}} \\
 & \multicolumn{2}{c}{\small{MC events}} & \multicolumn{2}{c}{\small{ }} & \multicolumn{2}{c}{\small{MC events}} & \multicolumn{2}{c}{\small{ }} & \multicolumn{2}{c}{\small{ }} & \multicolumn{2}{c}{\small{cuts}} \\

\hline
$e^-$/$e^+$ polarization & \small{eLpR} & \small{eRpL} & \small{eLpR} & \small{eRpL} & \small{eLpR} & \small{eRpL} & \small{eLpR} & \small{eRpL} & \small{eLpR} & \small{eRpL} & \small{eLpR} & \small{eRpL}\\
\hline

\small{$q\bar{q}$} & \small{$6.4\cdot10^8$} & \small{$3.5\cdot10^8$} & \small{0.23} & \small{$1.4\cdot10^{-2}$} & \small{$3.1\cdot10^{7}$} & \small{$1.1\cdot10^{6}$} & \small{$1.3\cdot10^{7}$} & \small{$1.1\cdot10^{6}$} & \small{$6.0\cdot10^{6}$} & \small{$2.1\cdot10^{5}$} & \small{39.0} & \small{1.0} \\

\small{$W(q\bar{q})W(q\bar{q})$} & \small{7.0$\cdot$10$^7$} & \small{7.0$\cdot$10$^5$} & \small{0.23} & \small{$1.3\cdot10^{-2}$} & \small{1.6$\cdot$10$^7$} & \small{1.6$\cdot$10$^5$} & \small{1.5$\cdot$10$^6$} & \small{809.3} & \small{$1.1\cdot10^{4}$} & \small{6.5} & \small{2.0} & \small{0.0}\\

\small{$Z(q\bar{q})Z(q\bar{q})$} & \small{7.0$\cdot$10$^6$} & \small{3.0$\cdot$10$^6$} & \small{0.23} & \small{$1.4\cdot10^{-2}$} & \small{1.6$\cdot$10$^6$} & \small{4.2$\cdot$10$^4$} & \small{1.4$\cdot$10$^5$} & \small{3.7$\cdot$10$^3$} & \small{5.0$\cdot$10$^4$} & \small{1.4$\cdot$10$^3$} & \small{12.0} & \small{0.0}\\

\small{$Z/W(q\bar{q})$}\small{$Z/W(q\bar{q})$} & \small{6.0$\cdot$10$^7$} & \small{10$^6$} & \small{0.23} & \small{$1.3\cdot10^{-2}$} & \small{1.4$\cdot$10$^7$} & \small{1.3$\cdot$10$^4$} & \small{$1.2\cdot10^{6}$} & \small{1.4$\cdot$10$^3$} & \small{8.3$\cdot$10$^3$} & \small{10.0} & \small{2.0} & \small{0.0}\\

\small{$Z(q\bar{q})Z(\mu\mu$}\small{$/\tau\tau)$} & \small{4.0$\cdot$10$^6$} & \small{2.0$\cdot$10$^6$} & \small{0.21} & \small{$1.1\cdot10^{-2}$} & \small{$8.4\cdot10^{5}$} & \small{$2.2\cdot10^{4}$} & \small{$7.2\cdot10^{4}$} & \small{$2.3\cdot10^{3}$} & \small{$1.2\cdot10^{4}$} & \small{373.7} & \small{14.0} & \small{0.0}\\

\small{$q\bar{q}H(b\bar{b})$} & \small{$10^5$} & \small{$10^5$} & \small{0.17} & \small{$6.6\cdot10^{-3}$} & \small{1.7$\cdot$10$^4$} & \small{657.6} & \small{961.2} & \small{38.1} & \small{860.7} & \small{33.9} & \small{1.0} & \small{0.0}\\

\small{$q\bar{q}H(\tau^+\tau^-)$} & \small{5.0$\cdot$10$^5$} & \small{5.0$\cdot$10$^5$} & \small{$0.04$} & \small{$1.5\cdot10^{-3}$} & \small{2.0$\cdot$10$^4$} & \small{760.9} & \small{2.1$\cdot$10$^3$} & \small{83.4} & \small{403.3} & \small{15.3} & \small{1.0} & \small{0.0}\\

\small{$q\bar{q}H(W^+W^-)$} & \small{5.0$\cdot$10$^5$} & \small{5.0$\cdot$10$^5$} & \small{0.14} & \small{$3.0\cdot10^{-3}$} & \small{7$\cdot$10$^4$} & \small{1.5$\cdot$10$^3$} & \small{3.5$\cdot$10$^3$} & \small{77.1} & \small{623.1} & \small{13.6} & \small{1.0} & \small{0.0}\\

\small{$q\bar{q}H(ZZ)$} & \small{5.0$\cdot$10$^5$} & \small{5.0$\cdot$10$^5$} & \small{0.16} & \small{6$\cdot$10$^{-3}$} & \small{$7.9\cdot10^{4}$} & \small{$3\cdot10^{3}$} & \small{4.9$\cdot$10$^3$} & \small{187.2} & \small{1.6$\cdot$10$^3$} & \small{62.2} & \small{2.0} & \small{1.0}\\

\small{$\tau^+\tau^-H($all$)$} & \small{5.0$\cdot$10$^5$} & \small{5.0$\cdot$10$^5$} & \small{0.12} & \small{4.4$\cdot$10$^{-3}$} & \small{5.8$\cdot$10$^4$} & \small{2.2$\cdot$10$^{3}$} & \small{5.5$\cdot$10$^{3}$} & \small{206.0} & \small{2.9$\cdot$10$^3$} & \small{108.7} & \small{13.0} & \small{0.0}\\
\hline
\hline
\end{tabular}
\end{center}
\end{table*}

\begin{figure}[!ht]\label{fig:MD} 
\centering
\includegraphics[scale=0.4]{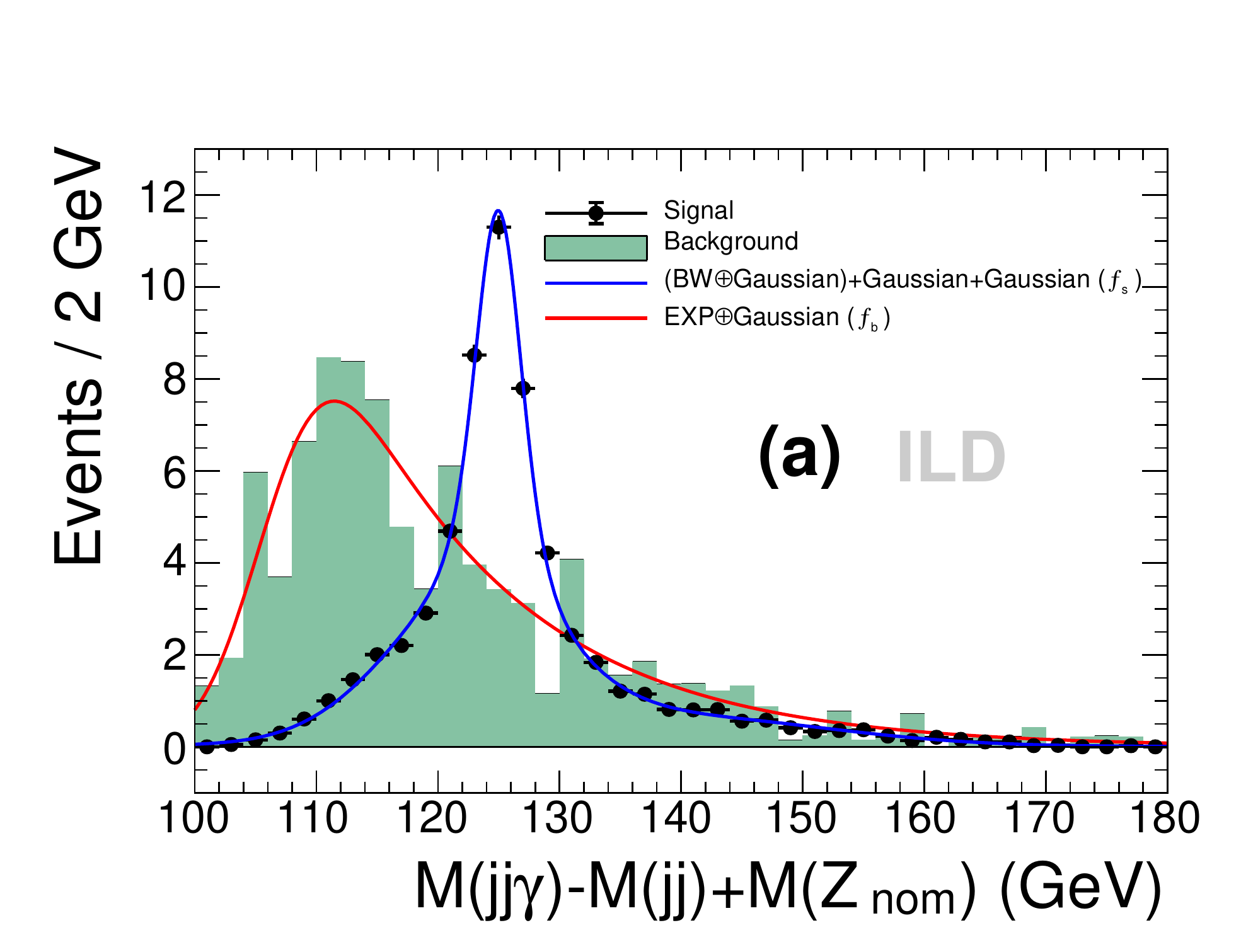}\vspace{-0.2cm}
\includegraphics[scale=0.4]{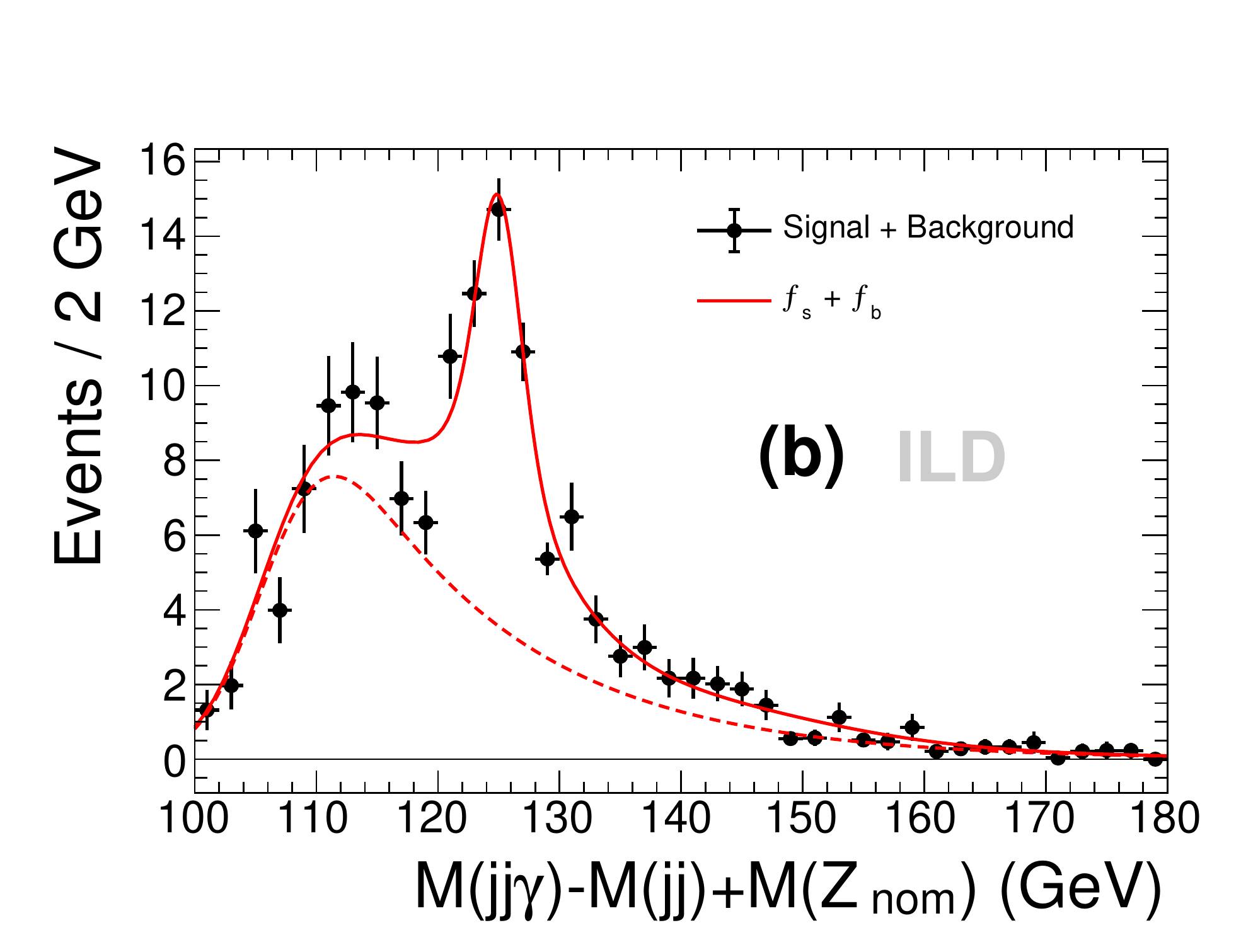}
\caption{The $M_{\Delta} = M(jj\gamma) - M(jj) + M(Z_{nom})$ mass distributions are shown for the $e^+e^- \to Z_1(j_1 j_2) \, H(Z \gamma)$ process followed by the decays $Z \to j_3 j_4$. (a)~The distributions are presented separately for the signal (full dots) and background (shaded histogram). The fit results are overlaid: a blue solid curve for the signal and a red dashed curve for background. (b) The sum of the signal and background contributions is shown by full dots together with the fit results: red dashed curve for background and the red solid curve for the sum. The functions and the fit methods are described in the text.}
\end{figure}

The signal and background $M_{\Delta}$ distributions after all cuts are fitted to obtain shape parameters separately for the signal and background (\hyperref[fig:MD]{Fig. 4a}). The error bars indicate the MC statistical uncertainties, which are much smaller than the fluctuations of the expected data. The signal distribution $F_S(m)$ is modelled by the sum of three functions: a Breit-Wigner function BW convolved with a Gaussian function $G_1$ and two additional Gaussian functions $G_2$ and $G_3$ to account events due to a wrong jet matching in the ${\chi}^2$ selection in the both tails of the distribution:

\begin{equation}
F_S(m) =  f_1 \, {\rm BW}\otimes G_1  + (1-f_1) \times [ f_2 \,G_2 + (1-f_2) \,G_3 ]
\end{equation} 

The corresponding fractions are denoted as $f_1$ and $f_2$. The width of the Breit-Wigner function is fixed to the value
$\Gamma$ = 2.495~GeV, because the $Z$ boson natural width transfers into the $M_{\Delta}$ value. The mean value of the first Gaussian is fixed to zero.

The background is described by the so-called decay function $F_B(m)$, which is an exponential function convolved with a Gaussian function:

\begin{equation}
F_B(m) = \exp(-m/\tau) \otimes G_4
\end{equation}

The obtained signal and background fit parameters are given in Table~\ref{tab:tab4}.

\renewcommand{\arraystretch}{1.2}
\begin{table}[htb]
\caption{The parameters obtained from the separate signal and background fits shown in \hyperref[fig:MD]{Fig. 4a}.}
\begin{center}
\label{tab:tab4}
\begin{tabular}
{@{\hspace{0.0cm}}l@{\hspace{0.3cm}}c@{\hspace{0.5cm}}c@{\hspace{0.0cm}}}
\hline \hline
\multicolumn{2}{c} {Signal} \\
\hline
BW mean, $\mu$ & 124.99$\,\pm\,$0.06 GeV \\
$G_1$ width & 1.38$\,\pm\,$0.09 GeV \\
$G_2$ mean & 140.63$\,\pm\,$1.76 GeV \\
$G_2$ width & 12.05$\,\pm\,$0.75 GeV \\
$G_3$ mean & 122.54$\,\pm\,$0.31 GeV \\
$G_3$ width & 7.11$\,\pm\,$0.18 GeV \\
Fraction $f_1$ & 0.55$\,\pm\,$0.02 \\
Fraction $f_2$ & 0.73$\,\pm\,$0.04 \\
\hline
\hline
\multicolumn{2}{c} {Background} \\
\hline
Exponential $\tau$ & 14.59$\,\pm\,$0.97 \\
$G_4$ mean & 106.08$\,\pm\,$0.59 GeV \\
$G_4$ width & 4.18$\,\pm\,$0.66 GeV \\
\hline
\hline
\end{tabular}
\end{center}
\end{table}

Then, the signal statistical uncertainties are estimated using the obtained distribution shapes and normalizations. To reproduce the real data distribution, the weighted signal and background distributions are summed, the content of each bin is rounded to the integer number and the Poisson uncertainties for the bin contents are assumed. \hyperref[fig:MD]{Figure 4b} shows the $M_{\Delta}$ distribution for the sum of the signal and background events.

The distribution of the sum of the signal and background contributions is fitted with the sum of the functions
used in the separate fits with fixed shapes and free normalizations. The binned extended maximum likelihood fit method~\cite{roo} is applied to obtain the number of signal events. A clear signal peak is observed in the combined distribution. The fit yields 60$\,\pm\,$13 signal events and 89$\,\pm\,$14 background events. The signal number of events corresponds to the statistical uncertainty of 22$\%$. 

\begin{figure}[!ht]\label{fig:Toy} 
\centering
\includegraphics[scale=0.4]{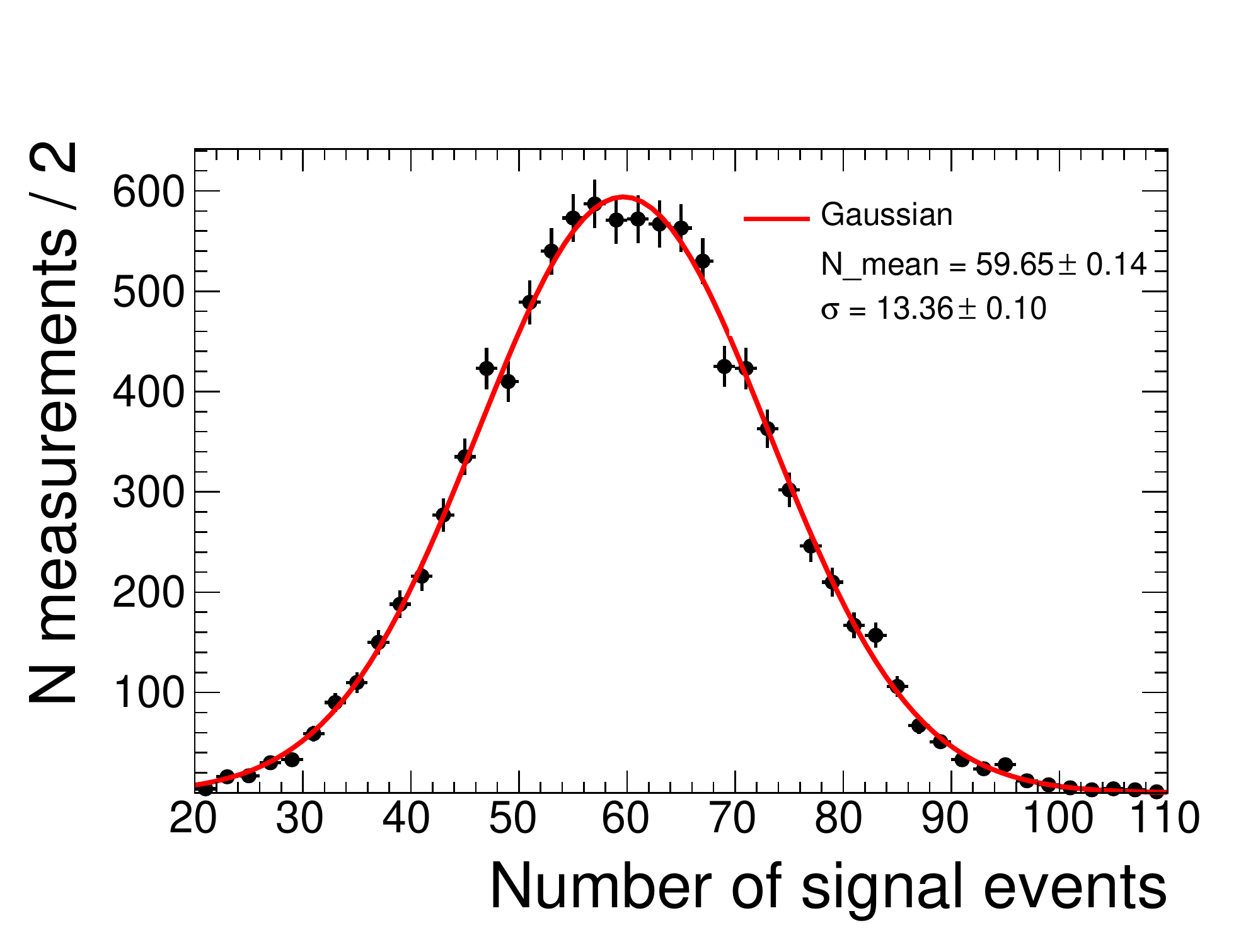}
\caption{The distribution of the number of the signal events obtained from the toy MC fits (dots with errors) is shown
together with a fit by the Gaussian function (curve) as described in the text.}
\end{figure}

The signal significance is checked with a toy MC using the RooFit package. 10000 $M_{\Delta}$ mass distributions are generated using the shapes and normalizations for the sum of the signal and background distributions obtained separately. The generated mass distributions are fitted with a function including both signal and background terms with free normalizations. \hyperref[fig:Toy]{Figure 5} shows the distribution of the numbers of the signal events obtained from the toy MC. The fit of this distribution to the Gaussian function gives the mean value and width of 60$\,\pm\,$13 events, respectively. The toy MC results agree within uncertainties with the combined fit results. Therefore the statistical uncertainty is 22$\%$ for an assumed dataset of 2 ab$^{-1}$ with ${\cal{P}}_{e^-e^+} = (-0.8, +0.3)$. The ILC strawman running scenario foresees two 0.9 ab$^{-1}$ datasets with ${\cal{P}}_{e^-e^+} = (-0.8, +0.3)$ and ${\cal{P}}_{e^-e^+} = \mbox{(+0.8, \-0.3)}$ each, plus 0.1 ab$^{-1}$ with ${\cal{P}}_{e^-e^+} = (-0.8, -0.3)$ and ${\cal{P}}_{e^-e^+} = (+0.8, +0.3)$ each~\cite{snow}. We therefore also reweighted the events passing our analysis to the combination of 0.9 ab$^{-1}$ with  ${\cal{P}}_{e^-e^+} = (-0.8, +0.3)$ and 0.9~ab$^{-1}$ with ${\cal{P}}_{e^-e^+} = (+0.8, -0.3)$, obtaining a statistical precision of 24$\,\%$. This result takes into account the increase in effective luminosity compared to an unpolarised dataset of 1.8 ab$^{-1}$, but not the full advantage of polarised beams. For optimal results, the selection should be tuned separately for each of the datasets, in order to exploit their different intrinsic signal-to-background ratios. We leave this part for future work.

The systematic uncertainties are not studied in this analysis. The largest systematic uncertainties are expected from the uncertainty in the selection efficiency and the uncertainty due to the signal and background shape modelling in the fit. The total systematic uncertainty is roughly evaluated to be less than 3$\,\%$, which is much smaller than the expected statistical one. Unfortunately accurate estimates of the systematic uncertainties cannot be performed without real data.

\section{\label{sec:conclusions} Conclusions}

Using MC method we simulated the $e^+e^- \to HZ$ process with subsequent $H \to Z \gamma$ decay as well as backgrounds in the ILD detector planned at the future ILC collider. The generation is performed assuming an integrated luminosity of 2 ab$^{-1}$, center-of-mass energy of 250 GeV, and beam polarizations of ${\cal{P}}_{e^-e^+} = (-0.8, +0.3)$. The statistical uncertainty of 22$\,\%$ is obtained for the number of the signal events. We also repeated the analysis assuming two data samples with integrated luminosities of 0.9 ab$^{-1}$ and two beam polarizations of ${\cal{P}}_{e^-e^+} = (\mp0.8, \pm0.3)$ and obtained the statistical uncertainty of 24$\,\%$. The accuracy of this method is about the same as one obtained at CEPC~\cite{cepc}, where the decay of one of the $Z$ bosons to the neutrino channel has been used to determine the branching fraction. Because the cross section $\sigma (e^+e^- \to HZ)$ can be determined with a high accuracy of better than 1$\,\%$ using other final states~\cite{eft}, the obtained uncertainties directly correspond to the uncertainty of the potential ${\cal B}r(H \to Z\gamma)$ measurement. A slightly better accuracy can be obtained with a future development of advanced event reconstruction technologies, such as a full kinematic fit, a multivariate analysis, and the specific $b$-jet treating taking into account secondary displaced vertices. The results of this method and of the method proposed in Ref.~\cite{cepc} can be combined to further improve the accuracy.

\hspace{0.1cm}
\section*{\label{sec:acknowledgements}ACKNOWLEDGMENTS}

Authors are grateful to I. Bozovic-Jelisavcic, Jenny List, Kiyotomo Kawagoe, Junping Tian, Daniel Jeans, Alberto Ruiz and Ties Behnke for useful discussions. We would like to thank the LCC generator working group and the ILD software working group for providing the simulation and reconstruction tools and producing the Monte Carlo samples used in this study. This work has benefited from computing services provided by the ILC Virtual Organization, supported by the national resource providers of the EGI Federation and the Open Science GRID. The work is supported by the Ministry of Science and Higher Education of the Russian Federation, Agreement No. 14.W03.31.0026.


\begin{thebibliography}{99}

\bibitem{atlas} G.~Aad {\it et al.} (ATLAS Collaboration), Observation of a new particle in the search
for the Standard Model Higgs boson with the ATLAS detector at the LHC, \href{https://www.sciencedirect.com/science/article/pii/S037026931200857X}{Phys.\ Lett.\ B {\bf 716}, 1 (2012)}.

\bibitem{cms} S.~Chatrchyan {\it et al.} (CMS Collaboration), Observation of a new boson at a mass
of 125 GeV with the CMS experiment at the LHC, \href{https://www.sciencedirect.com/science/article/pii/S0370269312008581}{Phys.\ Lett.\ B {\bf 716}, 30 (2012)}.

\bibitem{atlasrun2} G.~Aad {\it et al.} (ATLAS), A search for the $Z\gamma$ decay mode of the Higgs boson in $pp$ collisions at $\sqrt{s}$ = 13 TeV with the ATLAS detector, \href{https://www.sciencedirect.com/science/article/pii/S0370269320305578?via\%3Dihub}{Phys. Lett. B 809, 135754 (2020)}, [\href{https://arxiv.org/abs/2005.05382}{arXiv:2005.05382}].

\bibitem{cms2022} CMS Collaboration, Search for Higgs boson decays to a $Z$ boson and a photon in proton-proton collisions at $\sqrt{s}$ = 13 TeV, \href{https://arxiv.org/abs/2204.12945}{arXiv:2204.12945}

\bibitem{lhc} ATLAS and CMS Collaborations, Snowmass White Paper Contribution: Physics with the Phase-2 ATLAS and CMS Detectors, \href{https://cds.cern.ch/record/2805993/files/ATL-PHYS-PUB-2022-018.pdf}{ATLAS-PHYS-PUB-2022-018 and CMS PAS-FTR-22-001}.

\bibitem{hzzm} P.~Bambade {\it et al.}, The international linear collider: A global project, \href{https://arxiv.org/abs/1903.01629}{arXiv:1903.01629}.

\bibitem{wmass} L. D. Luzio, R. Gröber, P. Paradisi, Higgs physics confronts the $M_W$ anomaly, \href{https://arxiv.org/abs/2204.05284?context=hep-ex}{arXiv:2204.05284}.

\bibitem{cepc} Fenfen An {\it et al.}, Precision Higgs physics at the CEPC, \href{https://iopscience.iop.org/article/10.1088/1674-1137/43/4/043002}{Chinese Phys. C 43 043002 (2019)}.

\bibitem{flavor} Y. Radkhorrami, J. List, Conceptual aspects for the improvement of the reconstruction of $b$- and $c$-jets at $e^+e^-$ Higgs Factories with ParticleFlow detectors, \href{https://arxiv.org/abs/2105.08480}{arXiv:2105.08480}.

\bibitem{whizard} W.~Kilian, T.~Ohl and J.~Reuter, WHIZARD: Simulating multi-particle processes at LHC and ILC, \href{https://link.springer.com/article/10.1140/epjc/s10052-011-1742-y}{Eur. Phys. J. C {\bf 71}, 1742 (2011)}.

\bibitem{lcio} S.~Alpin, J.~Engels, F.~Gaede, N. A.~Graf, T.~Johnson, and J.~McCormick, LCIO: A persistency framework and event data model for HEP, \href{https://ieeexplore.ieee.org/document/6551478}{{\it 2012 IEEE Nuclear Science Symposium and Medical Imaging Conference (NSS/MIC 2012 (2012)}, pp. 2075–2079}.  

\bibitem{pythia6} T.~Sjostrand, S.~Mrenna, and P.~Skands, Pythia 6.4 physics and manual, \href{https://iopscience.iop.org/article/10.1088/1126-6708/2006/05/026}{J. High Energy Phys. 05 (2006) 026}.

\bibitem{ilcsoft} R.~Poeschl, Software Tools for ILC Detector Studies, eConf {\bf C0705302}, PLE104 (2007).

\bibitem{dd4hep} A.~Sailer, M. Frank, F. Gaede, D. Hynds, S. Lu, N. Nikiforou, M. Petric, R. Simoniello, and G. Voutsinas (CLICdp, ILD Collaboration), DD4Hep based event reconstruction, \href{https://iopscience.iop.org/article/10.1088/1742-6596/898/4/042017}{J. Phys. Conf. Ser. {\bf 898}, 042017 (2017)}.

\bibitem{marlin} F.~Gaede, Marlin and LCCD—Software tools for the ILC, \href{https://www.sciencedirect.com/science/article/abs/pii/S0168900205022643}{Nucl. Instrum. Methods Phys. Res., Sect. A {\bf 559}, 177 (2006)}.

\bibitem{mcp} MCParticle Class Reference, \href{http://lcio.desy.de/v01-07/doc/doxygen\_api/html/classEVENT\_1\_1MCParticle.html}{http://lcio.desy.de/v01-07/doc/doxygen\_api/html/classEVENT\_1\_1MCParticle.\newline~html}.

\bibitem{pfo} J.~Marshall and M.~Thomson, Pandora particle flow algorithm, in \textit{Proceedings of CHEF2013 - Calorimetry for the High Energy Frontier} (2013), pp. 305–315.

\bibitem{fj} M.~Cacciari, G.P.~Salam, and G.~Soyez, FastJet user manual, \href{https://link.springer.com/article/10.1140/epjc/s10052-012-1896-2}{Eur. Phys. J. C {\bf 72} 1896, (2012)}.

\bibitem{tmva} A.~Hoecker {\it et al.}, TMVA: Toolkit for Multivariate Data Analysis with ROOT, CERN Report No. 2007-007, 2007.

\bibitem{vlc} M.~Boronat, J.~Fuster, I.~Garcia, Ph.~Roloff, R.~Simoniello, and M.~Vos, Jet reconstruction at high-energy electron–positron colliders, \href{https://link.springer.com/article/10.1140/epjc/s10052-018-5594-6}{Eur. Phys. J. C {\bf 78}, 144 (2018)}.

\bibitem{lcfiplus} S. Catani, Y. L. Dokshitzer, M. Olsson, G. Turnock and B. Webber, New clustering algorithm for multi - jet cross-sections in
$e^+e^-$ annihilation, \href{https://www.sciencedirect.com/science/article/abs/pii/037026939190196W?via\%3Dihub}{Phys.Lett. B 269, 432–438, 1991}.

\bibitem{pdg} P.A.~Zyla {\it et al.} (Particle Data Group), The review of particle physics 2020, \href{https://academic.oup.com/ptep/article/2020/8/083C01/5891211}{Prog. Theor. Exp. Phys. {\bf 2020}, 083C01 (2020)}.

\bibitem{roo} W. Verkerke, D. Kirkby, \href{https://root.cern.ch/download/doc/RooFit_Users_Manual_2.91-33.pdf}{RooFit Users Manual v2.91}.

\bibitem{snow} A. Aryshev {\it et al.}, The International Linear Collider: Report to Snowmass 2021, \href{https://arxiv.org/abs/2203.07622}{arXiv:2203.07622}

\bibitem{eft} T.~Barklow, K.~Fujii, S.~Jung, R.~Karl, J.~List, T.~Ogawa, M.E.~Peskin, and J.~Tian, Improved formalism for precision Higgs coupling fits, \href{https://journals.aps.org/prd/abstract/10.1103/PhysRevD.97.053003}{Phys. Rev. D {\bf 97}, 053003 (2018)}.

\end{thebibliography}
\end{document}